\documentclass{aa}
\usepackage{textcomp}
\usepackage{textcomp}

\usepackage[varg]{txfonts}
\usepackage{xcolor}
\usepackage[colorlinks=true,linkcolor=blue,citecolor=blue,urlcolor=blue]{hyperref}

\bibpunct{(}{)}{;}{a}{}{,}

\begin{document}

\title{The TNG50-SKIRT Atlas:
Multi-wavelength properties of galaxy discs in the TNG50 simulation}

\titlerunning{TNG50-SKIRT Atlas}

\author{%
Abdissa~Tassama~Emana\inst{\ref{Jimma},\ref{UGent}}\thanks{\email{abdissatassama.emana@ugent.be}}
\and
Maarten~Baes\inst{\ref{UGent}}
\and
Angelos~Nersesian\inst{\ref{UGent},\ref{ULiege}}
\and
Tolu~Biressa\inst{\ref{Jimma}}
\and
Andrea~Gebek\inst{\ref{Bath}}
\and
Marco~Martorano\inst{\ref{UGent}}
\and
Viviana Casasola\inst{\ref{Bologna}}
\and
Sena~Bokona~Tulu\inst{\ref{Jimma},\ref{UGent}}
}

\institute{%
Physics Department, College of Natural Sciences, Jimma University, PO Box 378, Jimma, Ethiopia
\label{Jimma}
\and
Department of Physics and Astronomy, Universiteit Gent, Proeftuinstraat 86 N3, B-9000 Gent, Belgium
\label{UGent}
\and
STAR Institute, Quartier Agora -- All\'ee du six Ao\^{u}t 19c, B-4000 Li\`ege, Belgium
\label{ULiege}
\and
INAF--Istituto di Radioastronomia, Via P. Gobetti 101, I-40129 Bologna, Italy
\label{Bologna}
\and
Department of Physics, University of Bath, Claverton Down, Bath, BA2 7AY, UK.
\label{Bath}
}

\date{\today}

\abstract
{}
{We investigate the wavelength dependence of disc scale lengths in galaxies from the TNG50–SKIRT Atlas (TSA), and assess whether the simulated galaxies reproduce the structural properties and scaling relations observed in real galaxies, as well as the physical origin of the wavelength dependence.}
{We measure disc scale lengths from one-dimensional radial surface-brightness profiles of multi-wavelength synthetic images. The disc-dominated region is defined interactively per galaxy and applied consistently across bands. We restrict the analysis to nearly face-on, star-forming galaxies and retain only systems whose {\textit{r}}-band profiles are well described by an exponential model. The resulting scale lengths are compared to observational samples and analysed using both dust-attenuated and dust-free images, as well as intrinsic stellar population maps.}
{We find that the majority of TSA galaxies exhibit exponential discs in realistic mock observations. TSA galaxies reproduce the observed disc scale length--stellar mass (luminosity) relation, but are systematically larger at fixed mass or luminosity. The dependence of disc scale length on morphology is weak. Disc scale lengths correlate strongly across wavelengths, but decrease systematically with increasing wavelength. When normalised by the stellar-mass scale length, the TSA galaxies exhibit a clear and monotonic trend consistent with observations over the optical-to-near-infrared range. Comparing dust-free and dust-attenuated images and linking the results to intrinsic galaxy properties, we find that stellar population gradients are the primary driver of this wavelength dependence. In particular, radial gradients in stellar age dominate the effect, while metallicity gradients play a secondary role. Dust attenuation further enhances the wavelength dependence at short wavelengths but is not its main origin.}
{The TSA provides a realistic framework for studying the multi-wavelength structure of galaxies. The simulated galaxies reproduce the observed wavelength dependence of disc scale length and show that this behaviour is primarily driven by stellar age gradients, with metallicity and dust playing subdominant roles. Residual offsets in absolute scale length indicate the remaining differences between simulated and observed galaxy structure.}

\keywords{galaxies: fundamental parameters -- galaxies: structure -- galaxies: morphology -- radiative transfer}

\maketitle
\nolinenumbers

\section{Introduction}
\label{Introduction.sec}

Galaxy sizes provide a fundamental diagnostic of galaxy formation and evolution. Size parameters such as effective radius correlate tightly with stellar mass, star formation activity, morphology, and environment, offering important constraints on the physical processes governing galaxy assembly \citep{Shen2003, Kauffmann2003, vanderWel2014, Lange2015, Graham2019, Martorano2023, Nersesian2023, Quilley2023, Ward2024}. For disc galaxies in particular, the radial distribution of light or mass encodes information on angular momentum build up, stellar population gradients, and dust attenuation, making disc structure a sensitive probe of both formation history and internal evolution \citep{Courteau2007, Graham2008, Fathi2010, van2011, SanchezBlazquez2014, Casasola2017, Wang2023}. Establishing whether cosmological simulations reproduce the observed structural properties of discs in realistic observational conditions is therefore an essential test of their physical realism.

Several metrics are commonly used to quantify galaxy size, including Petrosian radii, half-mass radii, and effective radii. While the effective radius is widely adopted because of its observational accessibility, it combines contributions from multiple structural components. Massive disc galaxies typically contain both a rotationally supported disc and a centrally concentrated bulge, so the half-light radius does not isolate the disc itself. A more physically meaningful descriptor of disc structure is the exponential scale length obtained by fitting an exponential function to the observed surface brightness profile over radii where the disc dominates \citep{Freeman1970}. This parameter directly traces the radial extent of the stellar disc and is therefore commonly used in observational studies as well as theoretical models. This makes the disc scale length a more direct and physically interpretable measure of disc structure.

A common assumption in galaxy formation models and bulge--disc decompositions is that stellar discs follow exponential surface brightness profiles. Observational studies, however, report a range of behaviours, including pure exponentials as well as broken or truncated profiles \citep[e.g.][]{Pohlen2006, MendezAbreu2017}. Reported fractions of single-exponential discs vary substantially between samples, likely reflecting differences in selection criteria and treatment of disc breaks. Cosmological simulations provide complementary insight \citep[e.g.][]{Dutton2009a, Dutton2009b}. \citet{Wang2023} showed that stellar mass surface density profiles in TNG50 galaxies are often close to exponential, although other components such as star formation rate (SFR) profiles may deviate more strongly. These results naturally raise the question of whether simulated discs remain exponential when viewed in realistic mock observations rather than in intrinsic stellar mass distributions.

Connecting simulations to observations requires accounting for the fact that observations measure light rather than mass. The appearance of a galaxy depends on stellar population gradients, dust attenuation, and projection effects, all of which can modify the observed surface-brightness profile. Radiative transfer post-processing is therefore essential for meaningful comparisons. The TNG50--SKIRT Atlas (TSA; \citealt{Baes2024a}) provides such a bridge by generating multi-wavelength synthetic images of TNG50 galaxies using full three-dimensional radiative transfer in realistic geometries. Because the intrinsic properties of each simulated galaxy are known, these models also allow the effects of dust to be isolated and quantified by comparing attenuated and dust-free images. This makes the atlas an ideal framework for testing whether simulated galaxies reproduce the observed structural properties of disc galaxies in a directly observable context.

Multi-wavelength observations show that disc sizes depend systematically on wavelength. Ultraviolet (UV) and optical bands tend to emphasise younger stellar populations that extend to larger radii, whereas near-infrared (NIR) light traces older, more centrally concentrated stars. Dust attenuation further modifies the observed profiles by preferentially suppressing short-wavelength emission in inner regions. As a result, both effective radii \citep[e.g.][]{Ko2005, LaBarbera2010, Kelvin2012, Vulcani2014, Kennedy2015, Kennedy2016, EuclidCollaborationQuilley2026} and disc scale lengths \citep[e.g.][]{deJong1996b, Gadotti2009, Fathi2010, Casasola2017} show systematic trends with wavelength. Although radiative transfer models predict such behaviour, the relative roles of stellar population gradients and dust attenuation remain debated \citep[e.g.][]{deJong1996b, MacArthur2003, Kelvin2012, Kennedy2016}. The controlled multi-wavelength modelling enabled by the TSA provides a powerful way to investigate both the existence and the physical origin of this trend.

Beyond their wavelength dependence, disc scale lengths also correlate strongly with global galaxy properties. Observational studies in the local Universe show tight relations between disc size and stellar mass, luminosity, and star-formation activity, with more massive and actively star-forming galaxies typically hosting more extended discs \citep{Courteau2007, Gadotti2009, Fathi2010, Casasola2017}. Testing whether simulated galaxies reproduce these scaling relations is an important validation of galaxy formation models and of the realism of their predicted internal structure.

In this paper, we use the TSA to investigate the observable structure of simulated disc galaxies. Our analysis is designed to address four main questions:
\begin{enumerate}
    \item Do TNG50 galaxies exhibit exponential stellar discs when analysed in realistic mock-observed surface-brightness profiles?
    \item Are the properties of these exponential discs consistent with those observed in real galaxies, in particular in their scaling relations with global galaxy properties?
    \item Do the simulated discs reproduce the observed wavelength dependence of disc scale lengths from the UV to the NIR?
    \item If such a wavelength dependence is present, is it primarily driven by stellar population gradients or by dust attenuation effects?
\end{enumerate}

This paper is organised as follows. In Sect.~{\ref{SampleMethods.sec}} we describe the simulated data and the methodology used to measure disc scale lengths. Sect.~{\ref{RbandScalelengths.sec}} and Sect.~{\ref{MultiwavelengthScalelengths.sec}} present the results on exponential disc structure and multi-wavelength scaling relations, including comparisons with observational samples. Sect.~{\ref{Discussion.sec}} discusses the implications of our findings, and Sect.~{\ref{Summary.sec}} summarises our conclusions.


\section{Sample and methods}
\label{SampleMethods.sec}

\subsection{The TNG50-SKIRT Atlas}
\label{TSA.ssec}

Our analysis is based on the TSA \citep{Baes2024a}, a library of synthetic multi-wavelength observations constructed from galaxies in the TNG50 cosmological magnetohydrodynamical simulation. TNG50 is the highest-resolution run of the IllustrisTNG project and follows the evolution of a periodic cubic volume of side length 51.7 comoving Mpc from high redshift to the present day \citep{Nelson2019b, Pillepich2019}. The simulation self-consistently models dark matter, gas, stars, magnetic fields, and supermassive black holes, and includes key physical processes such as gas cooling, star formation, stellar evolution, chemical enrichment, and feedback from supernovae and active galactic nuclei. With a baryonic mass resolution of $8.5\times10^4~{\text{M}}_\odot$ and a typical spatial resolution of $\sim70$--$140$~pc in star-forming regions, TNG50 resolves the internal structure of galaxies on sub-kpc scales while retaining a statistically significant galaxy population across a broad mass range.

The TSA is constructed from a stellar-mass-selected sample of 1154 galaxies at $z=0$, spanning $10^{9.8} \lesssim M_{\star}/{\text{M}}_\odot \lesssim 10^{12}$. For each galaxy, realistic synthetic observations are generated through 3D dust radiative transfer simulations performed with the Monte Carlo radiative transfer code \textsc{SKIRT} \citep{Camps2015a, Camps2020}. This forward-modelling approach converts the intrinsic simulation output into observable quantities by accounting for the emission of different stellar populations as well as absorption and scattering by interstellar dust in complex three-dimensional geometries. Such radiative transfer modelling is essential for meaningful comparisons between simulated galaxies and observations, as dust attenuation can significantly alter galaxy fluxes, colours, and apparent morphology \citep[e.g.][]{Byun1994, Mollenhoff2006, Graham2008, Gadotti2010, Pastrav2013a, Pastrav2013b, Trayford2017}.

For each galaxy, the atlas provides spatially resolved images in 22 broadband filters covering the UV to NIR wavelength range. The synthetic observations are generated for five viewing directions in order to sample orientation effects. In addition to the dust-attenuated images, corresponding dust-free images are produced using identical stellar emission models but without the dusty medium, allowing the impact of dust attenuation to be isolated directly. The atlas further includes maps of intrinsic physical quantities derived from the simulation data, such as stellar mass surface density, mean stellar age, mean stellar metallicity, and dust mass surface density. These maps share the same spatial resolution and projection as the synthetic images, enabling a direct and consistent comparison between observable and physical properties. All images and parameter maps have $1600\times1600$ pixels with a pixel scale of 100 pc, corresponding to a field of view of 160 kpc.

The combination of high spatial resolution, broad wavelength coverage, and matched intrinsic property maps makes the TSA particularly well suited for studies of galaxy structure and its wavelength dependence. Previous applications of the TSA include investigations of the distribution of galaxies in the UVJ diagram \citep{Baes2024a}, the wavelength dependence of galaxy effective radii \citep{Baes2024b}, the reconstruction of stellar population property maps \citep{EuclidCollaborationKovacic2025, EuclidCollaborationAbdurrouf2025} and star formation histories \citep{EuclidCollaborationNersesian2026}, and galaxy morphology through non-parametric indicators \citep{BokonaTulu2026}.

\subsection{Sample selection}
\label{SampleSelection.ssec}

We constructed our sample by selecting star-forming galaxies from the TNG50-SKIRT Atlas. Star-forming systems are identified using a specific star formation rate (sSFR) threshold of $\mathrm{sSFR} > 10^{-11}~\mathrm{yr}^{-1}$, a commonly adopted division between star-forming and quenched galaxies \citep{Brinchmann2004, Fontanot2009, Donnari2019, Paspaliaris2023}. This selection yields 869 star-forming galaxies in the TSA parent sample. We emphasise that this selection does not impose any morphological criteria: star-forming galaxies can contain substantial bulges and are not necessarily disc-dominated \citep{Morselli2017, Tacchella2019}. Our analysis instead focuses on identifying and measuring exponential disc components, irrespective of the overall bulge prominence. Consequently, our sample includes galaxies with a range of morphologies, as illustrated in Fig.~\ref{MorphologyDependence.fig}.

For each galaxy, the TSA provides synthetic observations for five different viewing directions. Two of these (O4 and O5) are antipodal and therefore correspond to nearly identical projected structures, apart from differences in dust attenuation and minor numerical effects. Following \citet{Baes2024b} and \citet{BokonaTulu2026}, we use only the four independent orientations O1--O4 for the statistical analysis.

To ensure reliable disc scale-length measurements, we restrict the analysis to nearly face-on projections, where radially averaged light profiles remain physically meaningful and projection effects are minimised. For each galaxy and orientation, the inclination is defined as the angle between the observer’s line of sight and the stellar angular momentum vector, computed from stellar particles located between 0.5 and two stellar half-mass radii. We retain only galaxy--observer combinations with inclinations $i < 30^\circ$. Each galaxy contributes at most one such projection to the final sample; galaxies for which none of the available orientations satisfies this criterion are excluded. After applying this selection, the final sample consists of 468 galaxy projections.

Neither our simulated sample nor the observational comparison samples (see Sect.~{\ref{ComparisonSamples.ssec}}) were selected based on the presence or absence of AGN activity. Direct AGN emission is not included in the SKIRT radiative transfer post-processing of the TNG50–SKIRT Atlas \citep{Baes2024a}, so our synthetic images are free from direct AGN emission by construction. Even if such emission were included, its contribution would be strongly concentrated in the central regions of galaxies, which are excluded from the radial intervals used to fit the exponential disc profiles (see Sect.~{\ref{Scalelengths.ssec}}). We therefore expect its impact on the measured disc scale lengths to be negligible. For the observational comparison samples, direct AGN contamination is also expected to be limited, although its impact depends on the adopted fitting methodology, and some residual contamination through extended PSF wings cannot be excluded. In particular, approximately half of the galaxies in the DustPedia comparison sample host low-luminosity AGN \citep{Casasola2017}, whose contribution to the fitted disc profiles is expected to be small. Importantly, AGN feedback is included in the TNG galaxy formation model and may affect the intrinsic structure and evolution of the simulated discs, independently of the absence of direct AGN emission in the synthetic images.

For each selected system, we analyse images in 10 broadband filters spanning the UV--NIR range: the LSST {\textit{ugriz}} bands, the 2MASS {\textit{JHK}}$_{\text{s}}$ bands, and the WISE W1 and W2 bands. This wavelength coverage enables a detailed investigation of the dependence of disc scale length on wavelength.

\subsection{Measurement of disc scale lengths}
\label{Scalelengths.ssec}

\begin{figure*}
\centering
\includegraphics[width=\textwidth]{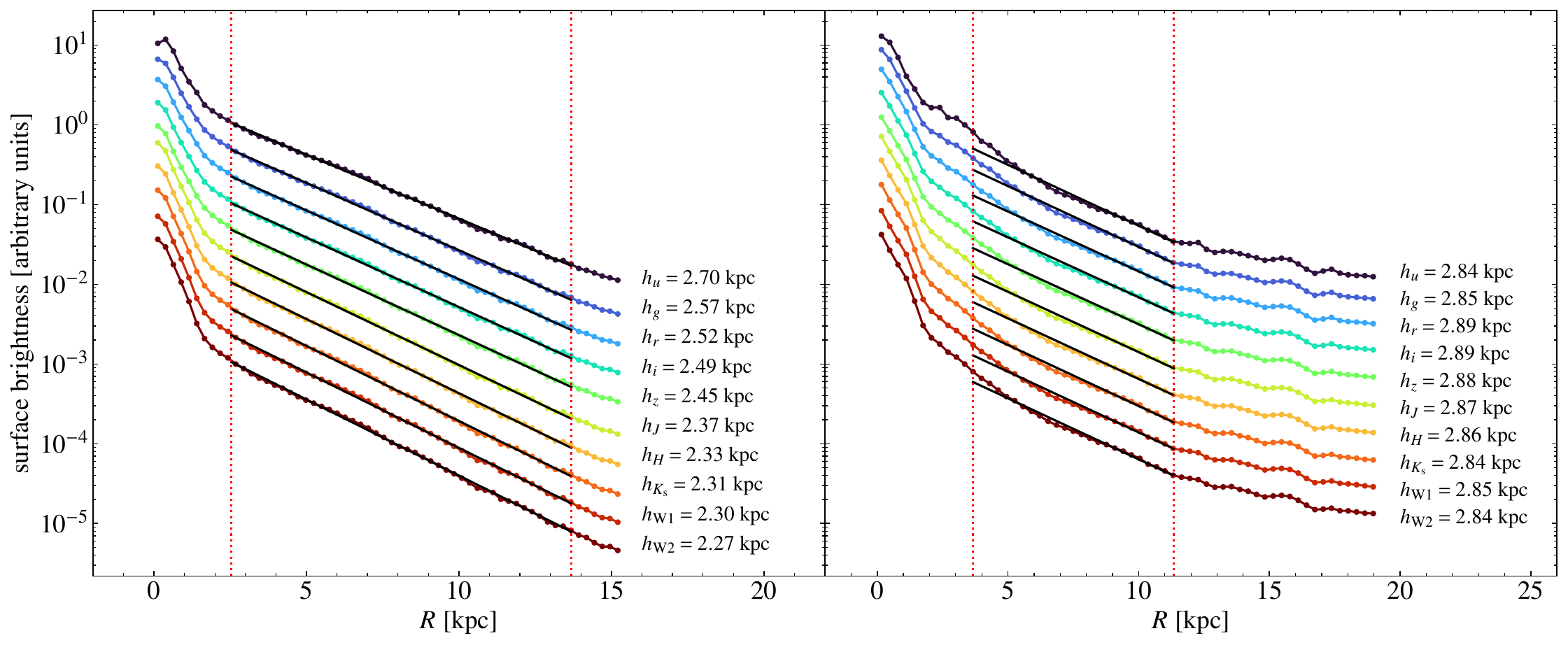} 
\caption{Multi-wavelength surface-brightness profiles for two representative galaxies (TNG063879\_O3) and (TNG096773\_O3), shown for all analysed bands. Coloured lines indicate the actual radial profiles, while solid black lines show the corresponding exponential fits. The vertical dotted lines mark the radial interval selected for the fit. The inferred disc scale length for each band is indicated. The left panel shows a representative good fit from the sample, while the right panel shows a poor fit that is excluded from the analysis.}
\label{ExampleProfiles.fig}
\end{figure*}

Disc scale lengths are derived from one-dimensional radial surface-brightness profiles. Although two-dimensional multi-component decompositions can in principle separate bulge, disc, and bar contributions in detail, their application to large samples requires substantial computational effort and model assumptions that are difficult to enforce uniformly. One-dimensional profile fitting provides a computationally efficient and widely adopted alternative for statistical studies of galaxy structure. Previous comparisons have shown that structural parameters obtained from 1D and 2D approaches are generally consistent, while 1D profiles can be particularly sensitive to the low-surface-brightness outskirts that strongly constrain disc scale lengths \citep{Gilhuly2018}. We therefore adopt a 1D approach for the present analysis.

For each galaxy projection, circularly averaged radial surface-brightness profiles are constructed using the \texttt{RadialProfile} functionality of the \texttt{photutils} package \citep{Bradley2025}. The profiles are centred on the galaxy centre and sampled in 60 linearly spaced radial bins defined in pixel units, extending from the centre to six stellar half-mass radii, ensuring that the full disc extent is covered while avoiding excessively low-signal regions at very large radii. Circular averaging is appropriate given that the sample is restricted to nearly face-on galaxies. Surface brightness is expressed in units of mag\,arcsec$^{-2}$. Pixel-level surface-brightness uncertainties are estimated as the standard deviation of intensities within a local $3\times3$ pixel neighbourhood, providing an empirical measure of local surface-brightness fluctuations. These uncertainties are propagated when computing the mean surface brightness in each radial bin and are used as weights in the fitting procedure.

Disc scale lengths are obtained by fitting an exponential surface-brightness model,
\begin{equation}
\mu_\lambda(R)=\mu_{\lambda,0}+1.086\left(\frac{R}{h_\lambda}\right),
\end{equation}
where $\mu_\lambda(R)$ is the surface brightness at wavelength $\lambda$ at radius $R$, $\mu_{\lambda,0}$ is the central surface brightness, and $h_\lambda$ the disc scale length. Since the fit is restricted to the radial interval where the disc dominates the light distribution, $\mu_{\lambda,0}$ represents the extrapolated central surface brightness of the exponential disc component, obtained by evaluating the best-fitting model at $R=0$. This is not necessarily equal to the actual surface brightness at the galaxy centre, which may contain additional contributions from a bulge or bar. Fits are performed using weighted non-linear least-squares minimisation.

Identifying the radial interval in which the exponential disc dominates the light distribution is non-trivial, since central regions may be affected by bulges or bars, while outer regions can be noisy or exhibit profile breaks. Automated methods struggle to robustly identify this regime across a diverse galaxy sample, motivating a controlled interactive approach. Following \citet{Casasola2017}, five authors independently inspect the full set of surface-brightness profiles for each galaxy across all available bands using a dedicated graphical interface that displays the profiles simultaneously and allows interactive selection of inner and outer fitting radii. The assessors work independently, and the galaxy order is randomised to minimise potential bias.

For each galaxy, a single radial interval is selected and applied to all bands, ensuring that differences in scale length reflect genuine wavelength-dependent structural variations rather than changes in the fitted radial range. Exponential fits are then performed within this interval for each band. Each assessor records the selected range, the fitted scale lengths, and a quality flag $q$ based on the {\textit{r}}-band profile, which serves as the reference for disc identification. The quality flag ranges from $q=0$ (poor fit) to $q=3$ (excellent fit). To obtain a robust consensus value while limiting the impact of outliers, the final quality parameter $Q$ is defined as the 20\% trimmed mean, that is the highest and lowest flags are discarded and the remaining three are averaged. Galaxies with $Q\ge2$ are considered to be well described by an exponential disc and are retained for subsequent analysis. For these galaxies, the final disc scale lengths are computed as the median of the values obtained by the five assessors.

Figure~{\ref{ExampleProfiles.fig}} shows the multi-wavelength surface-brightness profiles for a representative galaxy, together with the corresponding exponential fits. The vertical dashed lines indicate the adopted fitting range. The figure illustrates both the consistency of the fitting procedure across all bands and the systematic variation of disc scale length with wavelength, which is the central focus of this work.

\subsection{Comparison samples}
\label{ComparisonSamples.ssec}

\begin{figure}[t]
\includegraphics[width=0.95\columnwidth]{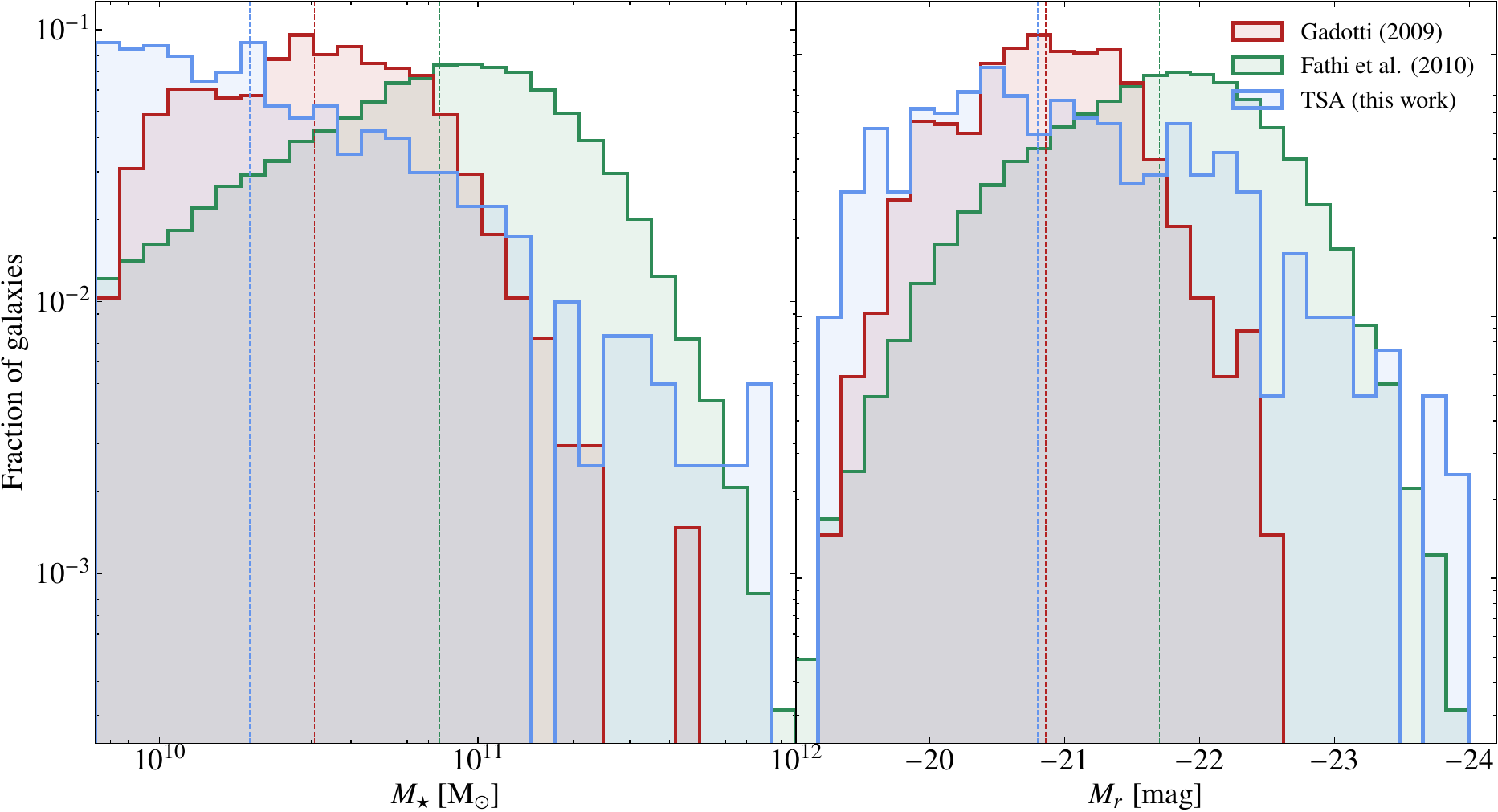}
\caption{Distribution of stellar mass (left) and absolute {\textit{r}}-band magnitude (right) for the TSA sample and the comparison samples of \citet{Gadotti2009} and \citet{Fathi2010}. Each distribution is normalised by the total number of galaxies in the
corresponding sample, so that the histograms show the fraction of galaxies per bin and samples of very different sizes can be compared directly. The dashed lines indicate the median values of each sample. While the samples cover a similar mass range, their distributions differ substantially, with the \citet{Gadotti2009} sample having a larger fraction of intermediate-mass galaxies and the \citet{Fathi2010} sample shifted towards higher stellar masses.}
\label{SampleHistograms.fig}
\end{figure}

To place our measurements in an observational context, we compare the TNG50 results with well-characterised observational samples that allow meaningful structural and multi-wavelength comparisons. In selecting suitable reference datasets, we required that the samples (i)~cover a stellar-mass range broadly comparable to that of the TSA galaxies, (ii)~provide independent determinations of disc scale lengths in multiple photometric bands, and (iii)~contain sufficiently large numbers of galaxies to enable robust statistical analysis. No single observational dataset simultaneously satisfies all these criteria. We therefore adopt three complementary comparison samples, each serving a distinct purpose in the analysis.

The first two samples are drawn from the Sloan Digital Sky Survey (SDSS) and are primarily used for statistical comparisons of absolute scale lengths and scaling relations in the optical regime (Sect.~{\ref{RbandScalelengths.sec}}). \citet{Gadotti2009} analysed a sample of nearly 1000 massive disc galaxies, selecting systems with stellar masses $M_\star > 10^{10}~{\text{M}}_\odot$ and restricting the sample to highly face-on galaxies ($b/a > 0.9$). Disc scale lengths were derived through detailed two-dimensional image decompositions using the \texttt{BUDDA} code \citep{deSouza2004}, modelling each galaxy with separate bulge, bar, and exponential disc components. Independent decompositions were performed in the SDSS {\textit{g}}, {\textit{r}}, and {\textit{i}} bands, making this dataset a well-controlled reference based on multi-component structural modelling.

\begin{figure*}[t]
\centering
\includegraphics[width=\textwidth]{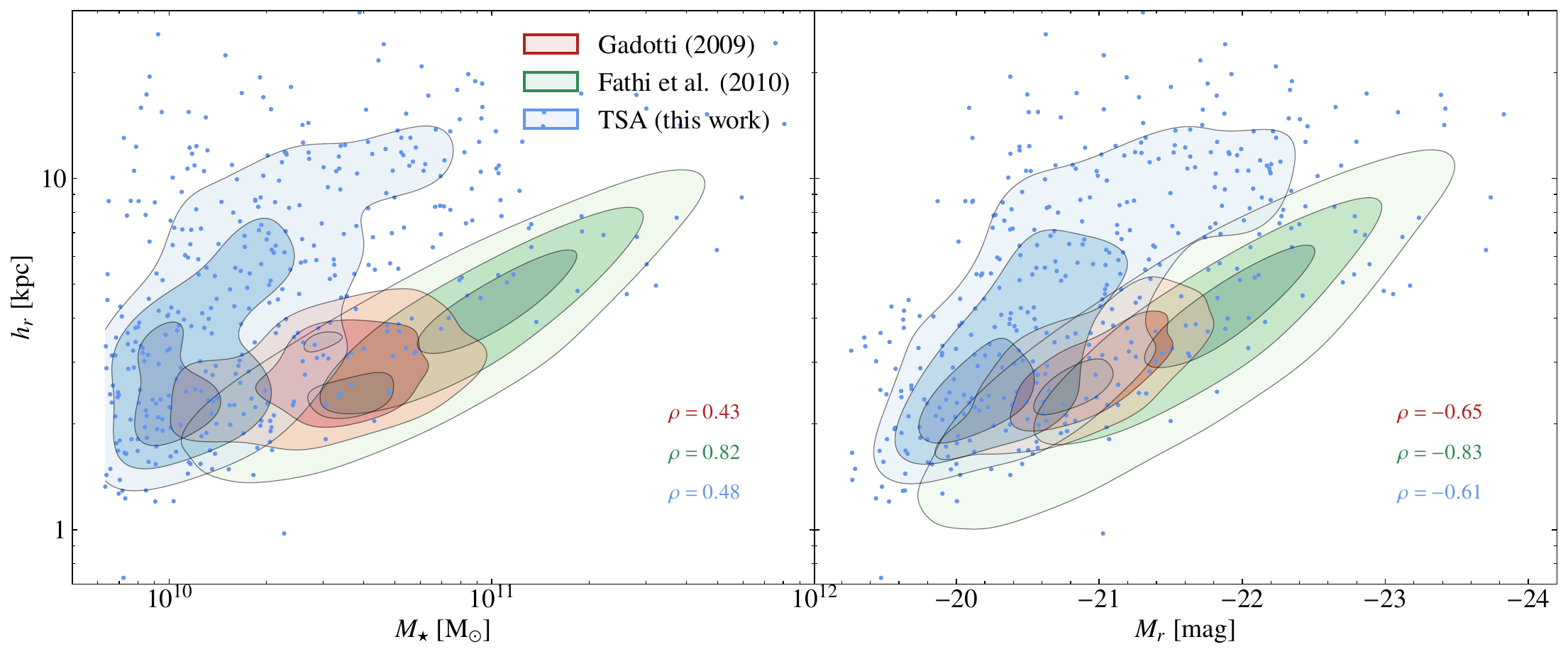}
\caption{Relation between {\textit{r}}-band disc scale length and stellar mass (left) and absolute {\textit{r}}-band magnitude (right) for TSA galaxies and the observational samples of \citet{Fathi2010} and \citet{Gadotti2009}. TSA galaxies are shown in blue, while the observational data are represented by coloured contours. The Spearman rank correlation coefficient ($\rho$) for each sample is reported in the bottom-right corner of each panel. The TSA galaxies exhibit qualitatively similar trends to the observational samples, although the correlations are considerably weaker than those reported by \citet{Fathi2010}. However, a systematic offset is evident over the full parameter range, with simulated galaxies exhibiting larger disc scale lengths at fixed stellar mass or luminosity.}
\label{ComparisonObservations.fig}
\end{figure*}

\citet{Fathi2010} compiled a substantially larger sample of approximately 30\,000 non-interacting disc galaxies from the HyperLEDA catalogue \citep{Paturel2003}, cross-matched with SDSS Data Release~7 \citep{Abazajian2009}. Their sample spans a wide range of spiral morphologies and includes disc scale lengths measured in all five SDSS $ugriz$ bands. In contrast to the multi-component decompositions of \citet{Gadotti2009}, their analysis relies on automated extraction of radial surface-brightness profiles and exponential fits, making it methodologically closer to the approach adopted in this work.

We restrict both SDSS-based samples to the same stellar-mass range as the TSA galaxies, $9.8 \le \log(M_\star/{\text{M}}_\odot) \le 12$. Despite this restriction, the stellar-mass and luminosity distributions of the observational samples differ significantly from that of the TSA galaxies, as shown in Fig.~{\ref{SampleHistograms.fig}}. The Kolmogorov–Smirnov (KS) tests confirm that the stellar-mass and luminosity distributions of the TSA and observational samples differ significantly. For the \citet{Gadotti2009} and \citet{Fathi2010} samples, respectively, the KS statistics are $D = 0.25$ and $0.47$ for stellar mass ($p < 10^{-13}$), and $D = 0.18$ and $0.36$ for $r$-band absolute magnitude ($p < 10^{-7}$). Both observational samples have higher median stellar masses than the TSA sample, by 0.20 and 0.60 dex, respectively. In addition, the \citet{Fathi2010} sample is brighter, with a median $r$-band absolute magnitude that is 0.9 mag lower. These differences in sample properties should be kept in mind when interpreting the comparisons of disc scale lengths and scaling relations.

For the analysis of the wavelength dependence of disc scale lengths over a broad spectral range, we additionally consider the study of \citet{Casasola2017}. They analysed a sample of 18 nearby, face-on spiral galaxies from the DustPedia survey \citep{Davies2017, Clark2018}, exploiting homogeneous multi-wavelength imaging from UV to submm. Although this sample is considerably smaller than the SDSS-based datasets and therefore less suited for population-level statistical comparisons, it offers extended wavelength coverage and provides stellar-mass scale lengths, making it particularly valuable for comparison with our analysis of the wavelength dependence of disc scale length (Sect.~{\ref{MultiwavelengthScalelengths.sec}}). Together, these three observational samples provide complementary benchmarks for assessing both the absolute structural properties of TNG50 discs and their wavelength-dependent behaviour.


\section{Results: {\it{r}}-band disc scale lengths}
\label{RbandScalelengths.sec}

In this section, we analyse {\textit{r}}-band disc scale lengths measured from the dust-attenuated SKIRT images. This choice provides a direct link to observational studies and allows us to address the first two questions of this work: whether TNG50 galaxies exhibit exponential discs in realistic mock observations, and whether their scale lengths are consistent with observed scaling relations. 

\subsection{Exponential disc quality in the {\textit{r}} band}
\label{Exponential.ssec}

To assess whether TNG50 galaxies exhibit exponential discs in realistic mock observations, we evaluate the quality of exponential fits to the {\textit{r}}-band surface-brightness profiles derived from the dust-attenuated images. The fitting methodology and selection of the disc-dominated radial range are described in Sect.~{\ref{Scalelengths.ssec}}. This approach provides a controlled way to identify galaxies whose radial light distributions are consistent with an exponential disc over a well-defined radial interval, without requiring full multi-component decompositions.

We find that the vast majority of galaxies in our sample (402 out of 468, or 86\%) satisfy the criterion $Q \ge 2$. This demonstrates that near-exponential discs are ubiquitous in the dust-attenuated {\textit{r}}-band mock observations of TNG50 galaxies. It is important to emphasise that this classification refers only to whether an exponential profile provides an adequate description of the disc-dominated region, and does not imply a classification of profile types (e.g.\ Type~I, II, or III discs). For the remainder of this work, we restrict the analysis to galaxies whose surface-brightness profiles are well described by an exponential model ($Q \ge 2$).

\subsection{Dependence on stellar mass and luminosity}

We now test whether the disc scale lengths of TNG50 galaxies are consistent with those observed in real galaxies by comparing them to the observational samples of \citet{Gadotti2009} and \citet{Fathi2010}. The left panel of Fig.~\ref{ComparisonObservations.fig} shows the relation between stellar mass and {\textit{r}}-band disc scale length. The TSA galaxies exhibit a moderate positive correlation between stellar mass and disc scale length, in qualitative agreement with the trends observed in both comparison samples. The Spearman rank correlation coefficient for the TSA galaxies ($\rho=0.48$) is similar to that found by \citet{Gadotti2009} ($\rho=0.43$), but considerably smaller than that reported by \citet{Fathi2010} ($\rho=0.82$). The differences in correlation strength between the samples may reflect differences in sample selection and methodology, particularly in the treatment of central galaxy components and inclination effects. Although our approach relies on one-dimensional radial profiles, as in \citet{Fathi2010}, our interactive selection of the disc-dominated fitting region aims to isolate the exponential disc component from central structures, similar in spirit to the multi-component decompositions of \citet{Gadotti2009}.

Furthermore, a systematic offset is present: at fixed stellar mass, the TSA galaxies exhibit larger disc scale lengths than observed galaxies. This offset persists across the full stellar-mass range probed, suggesting a global difference in the normalisation of the size--mass relation rather than a mass-dependent effect.

The right panel shows the corresponding relation between absolute {\textit{r}}-band magnitude and disc scale length. As in the stellar-mass relation, brighter galaxies host larger discs, and the TSA galaxies exhibit a qualitatively similar trend to that observed in the comparison samples. The same systematic offset is again present, with simulated galaxies appearing larger at fixed luminosity. We also note that the correlation in the TSA sample is somewhat stronger when considering absolute {\textit{r}}-band magnitude rather than stellar mass, with the absolute value of the Spearman coefficient increasing from 0.48 to 0.61.

These comparisons show that TNG50 reproduces the general trend of the observed disc size scaling relations, but not their normalisation. This finding is consistent with previous analyses of TNG50 galaxy structure \citep[e.g.][]{Wang2023}, and we discuss possible physical origins of this offset in Sect.~{\ref{Discussion.sec}}.

\subsection{Dependence on structure}
\label{structure.sec}
\begin{figure}[t]
\centering
\includegraphics[width=\columnwidth]{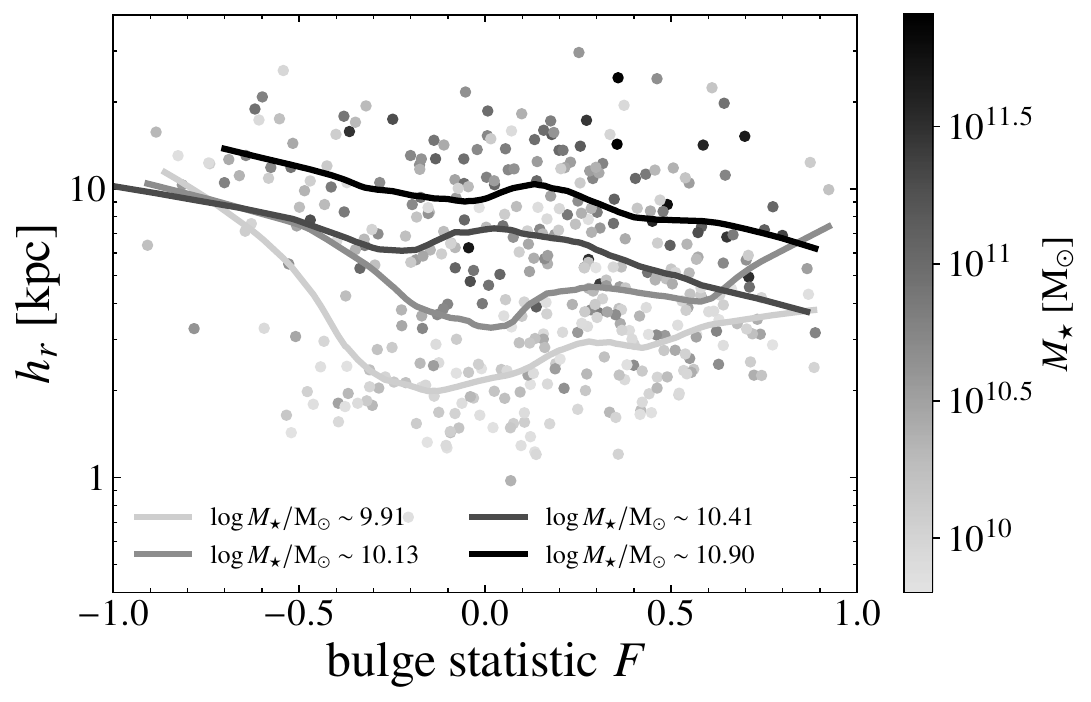}
\caption{{\textit{r}}-band disc scale length as a function of bulge statistic $F$ (a proxy for morphology) for TSA galaxies. Galaxies are colour-coded by stellar mass, as indicated. The curves show LOWESS trends in four stellar-mass bins containing equal numbers of galaxies. While disc scale length increases systematically with stellar mass, the LOWESS curves are nearly flat within each mass bin, indicating that the dependence of disc size on morphology is weak at fixed stellar mass.} 
\label{MorphologyDependence.fig}
\end{figure}

As an additional consistency check, we examine whether the {\textit{r}}-band disc scale length depends on galaxy morphology within the TNG50 sample. Observational studies have shown that disc scale lengths depend only weakly on Hubble type at fixed stellar mass or luminosity. For example, \citet{Fathi2010} found that {\textit{r}}-band scale lengths are largely independent of morphological type for well-classified spiral galaxies, consistent with earlier results by \citet{deJong1996a} and \citet{Graham2001}. Although the TSA does not provide visual Hubble classifications, it includes non-parametric morphology indicators that can serve as proxies for bulge prominence and thus for morphological stage.

We use the {\textit{r}}-band bulge statistic ($F$) introduced by \citet{Snyder2015}, which combines the Gini coefficient and the $M_{20}$ parameter. This quantity has been computed for all TSA images by \citet{BokonaTulu2026} using the \texttt{StatMorph} package \citep{RodriguezGomez2019}. Larger values correspond to more bulge-dominated systems, broadly analogous to earlier-type spirals, while lower values indicate more disc-dominated galaxies.

Figure~{\ref{MorphologyDependence.fig}} shows the {\textit{r}}-band disc scale length as a function of bulge statistic for our TSA galaxies, together with LOWESS trends\footnote{LOWESS (LOcally WEighted Scatterplot Smoothing) is a non-parametric technique that estimates a smooth trend by fitting simple models to local subsets of the data \citep{Cleveland1979}. It is used here for visual guidance only.} computed in four stellar-mass bins containing equal numbers of galaxies. The dominant dependence of disc scale length on stellar mass is clearly visible, with more massive galaxies hosting systematically larger discs. At fixed stellar mass, however, the LOWESS curves are nearly flat, indicating that the dependence of disc scale length on morphology is weak. The substantial overlap between bulge- and disc-dominated systems further supports the conclusion that morphology plays only a minor role in setting disc scale length compared to stellar mass.

Although this proxy-based analysis cannot be mapped directly onto Hubble-type classifications, the weak morphology dependence seen here is qualitatively consistent with observational findings \citep{deJong1996a, Fathi2010}. This supports the conclusion that the TSA galaxies reproduce not only the global size--mass relation of discs \citep{Genel2018,RodriguezGomez2019}, but also the observed weak coupling between disc scale length and morphological stage. A more detailed investigation of morphology-dependent structural trends in TNG50 is beyond the scope of this work.

\section{Multi-wavelength disc scale lengths}
\label{MultiwavelengthScalelengths.sec}

\subsection{Overview of multi-band disc scale lengths}

\begin{figure*}
\centering
\includegraphics[width=\textwidth]{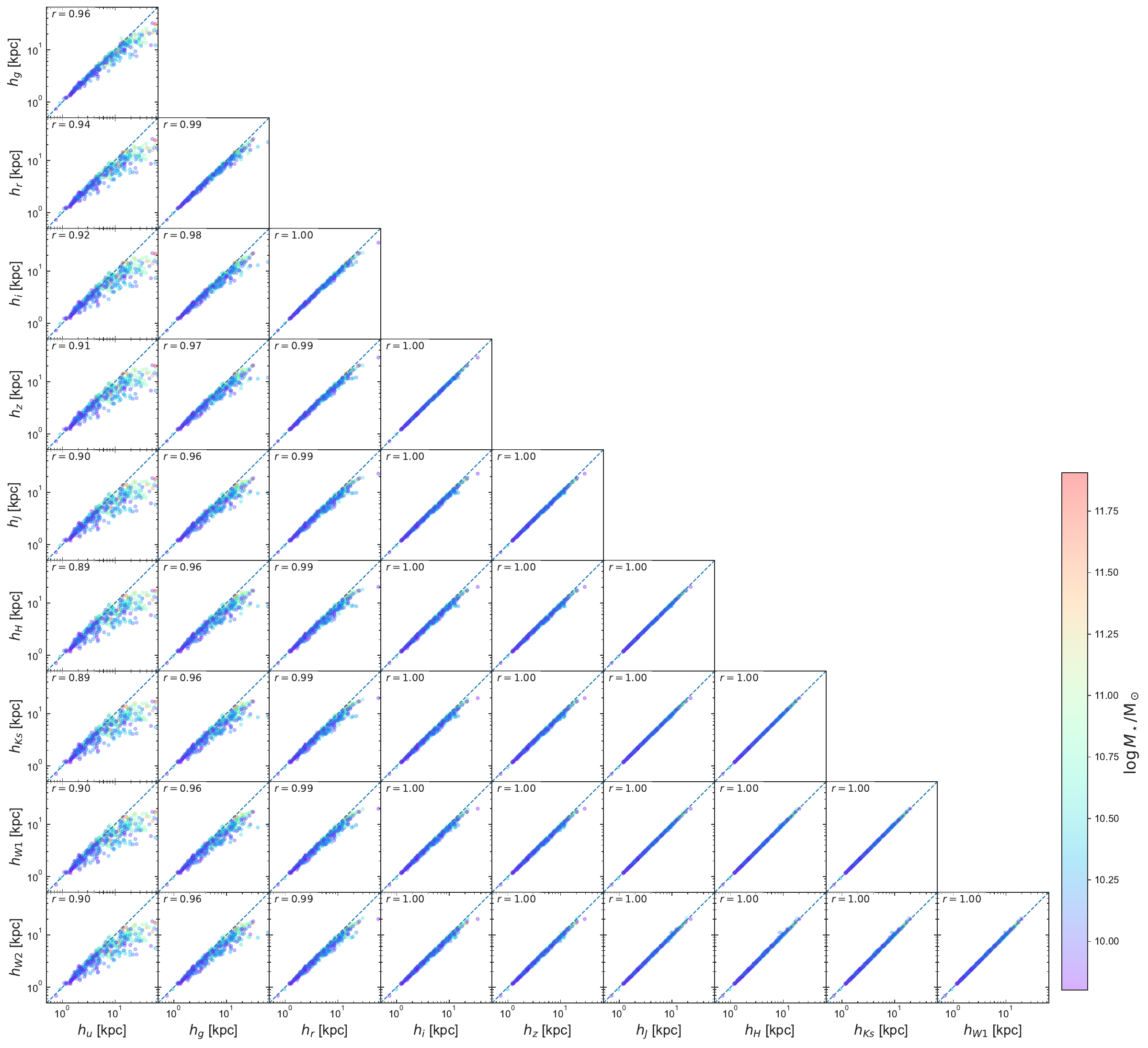}
\caption{Pairwise comparison of disc scale lengths measured in different photometric bands for the dust-attenuated TSA images. Each panel shows the scale length in one band plotted against that in another band for the same set of galaxies. Points are colour-coded by stellar mass. The diagonal line in each panel indicates the one-to-one relation. Scale lengths in different bands are strongly correlated, demonstrating that the exponential disc structure is largely preserved across wavelength. The increasing scatter and systematic offsets for band pairs with larger wavelength separation reflect the wavelength dependence of disc scale length, with shorter wavelengths generally yielding larger scale lengths. This is supported by the Pearson correlation analysis, which shows near-unity correlations for closely spaced bands and slightly reduced for widely separated bands (\textit{r}$\approx$0.90), but remaining strong.} 
\label{Pairwise.fig}
\end{figure*}

In this section, we investigate how the exponential disc scale length depends on wavelength using the multi-wavelength imaging products of the TSA. Observational studies have long reported systematic variations of disc scale length with bandpass, typically finding more extended discs at shorter wavelengths, and have interpreted these trends in terms of stellar population gradients and dust attenuation \citep{deJong1996a, Mollenhoff2006, Fathi2010, Casasola2022}. Before quantifying these effects and examining their physical origin, we first establish empirically how scale length measurements in different bands are related within our simulated sample.

Figure~{\ref{Pairwise.fig}} presents a pairwise comparison of disc scale lengths measured in all bands for the dust-attenuated images. For each galaxy, the same disc-dominated radial range is adopted in all bands, ensuring that any differences reflect genuine variations in the surface-brightness profiles rather than differences in the fitting procedure.

Several general features are immediately apparent. First, scale lengths measured in different bands are tightly correlated, indicating that the exponential disc structure is largely preserved across wavelength. This confirms the results reported by \citet{Fathi2010}. Second, the scatter about the one-to-one relation increases with increasing wavelength separation between bands: comparisons involving UV and NIR bands exhibit significantly larger dispersion than those between adjacent optical bands. Third, systematic offsets are present, such that scale lengths measured at shorter wavelengths tend to be larger than those measured at longer wavelengths. These trends suggest that the disc structure remains broadly coherent across wavelengths at the level of the observed light distribution, despite tracing different stellar populations and different levels of attenuation. Furthermore, the radial extent of the disc structure appears to vary systematically with the stellar populations and attenuation processes traced by different bands. 

The purpose of Fig.~{\ref{Pairwise.fig}} is to provide a compact overview of this multi-wavelength behaviour. In the following subsections, we quantify these trends using a physically motivated diagnostic based on the ratio of the scale length in each band to the corresponding stellar-mass scale length. We then examine how this relation depends on global galaxy properties and use the availability of dust-free images to disentangle the relative roles of stellar population gradients and dust attenuation.

\subsection{Wavelength dependence relative to the stellar-mass scale length}
\label{RelativeScalelength.ssec}

\begin{figure}[t]
\includegraphics[width=0.95\columnwidth]{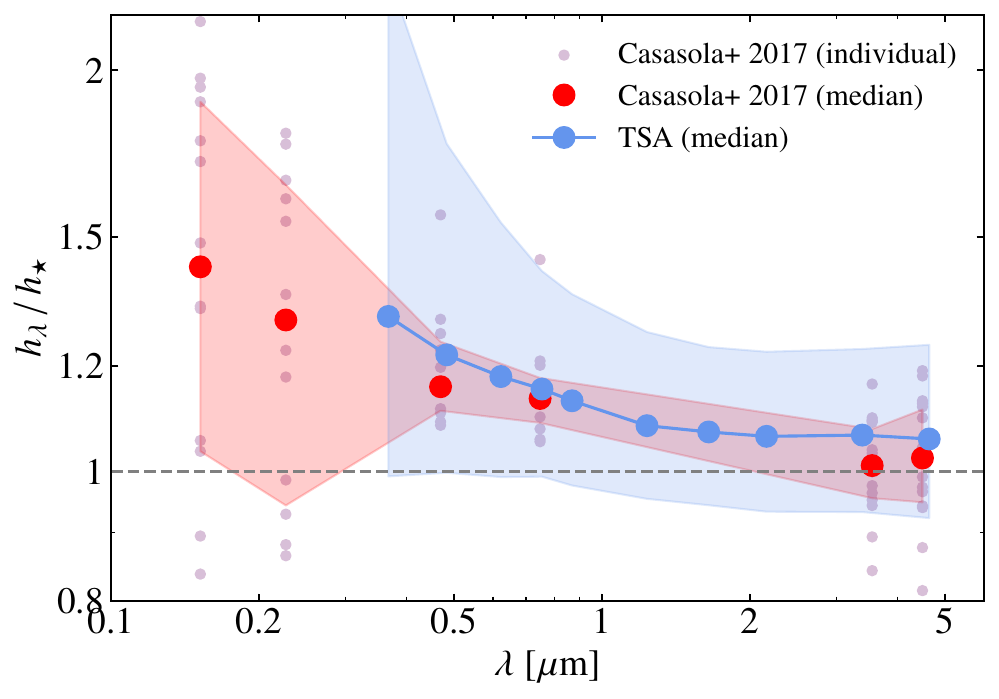}
\caption{Wavelength dependence of the disc scale length relative to the stellar-mass scale length, $h_\lambda/h_\star$, as a function of wavelength. Blue circles show the median relation for the TSA galaxies. Red circles indicate the median values for the sample of \citet{Casasola2017}, while faint purple points represent individual galaxies. The shaded regions mark the interquartile range (25th--75th percentile) of each sample at each wavelength. The \citet{Casasola2017} data extend to the FUV, whereas the TSA measurements are limited to the LSST {\textit{u}} band at the blue end. Both datasets exhibit a monotonic decrease of $h_\lambda/h_\star$ from the UV to the NIR, approaching unity at long wavelengths, with the TSA median relation lying within the observed scatter over the overlapping wavelength range.}
\label{ComparisonCasasola.fig}
\end{figure}

We quantify the wavelength dependence of disc scale lengths using the dimensionless ratio $h_\lambda/h_\star$, where $h_\lambda$ is the exponential scale length measured in band $\lambda$ and $h_\star$ is the scale length derived from the stellar surface-density profile. For the TSA galaxies, $\Sigma_\star(R)$ is obtained by projecting stellar particle masses onto the plane of the sky and fitting an exponential profile over the same disc-dominated radial range as used for the photometric bands. For the observational sample of \citet{Casasola2017}, stellar mass surface-density maps are constructed from NIR imaging using the IRAC-based prescription of \citet{Querejeta2015}, and exponential scale lengths are derived in an analogous manner. Although the construction of stellar mass maps differs between the two datasets, the definition of $h_\star$ as the exponential scale length of $\Sigma_\star(R)$ is conceptually consistent.

Figure~{\ref{ComparisonCasasola.fig}} shows $h_\lambda/h_\star$ as a function of wavelength for the TSA galaxies and for the sample of \citet{Casasola2017}. The TSA results (blue circles) represent median values over the full sample, while the \citet{Casasola2017} data are shown both for individual galaxies (faint purple points) and for their median relation (red circles). The \citet{Casasola2017} sample extends to far-UV (FUV) wavelengths, whereas the TSA sample only reaches blue wavelengths as in the {\textit{u}} band.

Both datasets show a clear and monotonic decrease in $h_\lambda/h_\star$ with increasing wavelength. In the UV and blue optical bands, the light distribution is more extended than the underlying stellar mass distribution, with $h_\lambda/h_\star > 1$. Toward longer wavelengths, the ratio approaches unity, indicating that NIR emission traces the stellar mass distribution more closely. The TSA median relation follows the observational trend and lies within the dispersion of the \citet{Casasola2017} galaxy sample across the shared wavelength range.

The main difference between the two samples is the larger scatter in the UV for the \citet{Casasola2017} galaxies, particularly at FUV wavelengths not covered by the TSA. This likely reflects a combination of limited sample size, variations in star–dust geometry, and the strong sensitivity of UV emission to recent star formation. Across the optical-to-NIR range, common to both datasets, the agreement is very good. This comparison shows that the TSA galaxies reproduce both the amplitude and the wavelength dependence of the observed relation between disc scale length and stellar mass distribution, providing further support for the realism of their resolved structure.

\subsection{Dependence on global galaxy properties}

\begin{figure*}[t]
\includegraphics[width=\textwidth]{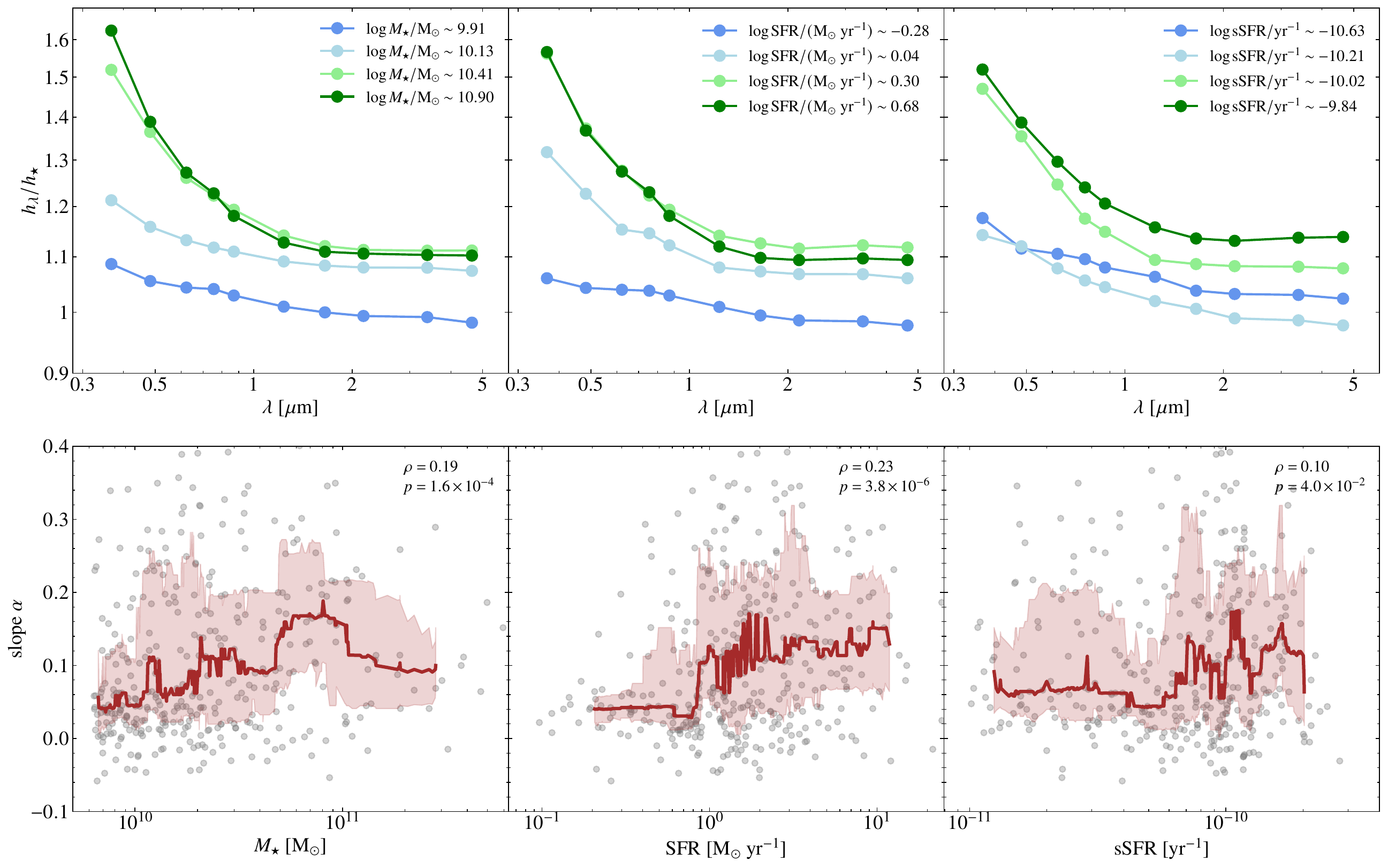}
\caption{ Dependence of the wavelength variation of disc scale length on global galaxy properties. \textit{Top panels:} Median relations between the relative disc scale length $h_\lambda/h_\star$ and wavelength for galaxies binned by $M_\star$ (left), SFR (middle), and sSFR (right). In all cases, shorter wavelengths correspond to more extended emission, with a modest trend for galaxies with higher $M_\star$ and SFR to exhibit a stronger wavelength dependence. \textit{Bottom panels:} Slope $\alpha$ of the relation $h_\lambda/h_\star \propto \lambda^{-\alpha}$ as a function of the same global properties. Individual galaxies are shown as grey points, while the red curves indicate the running median, with the shaded regions denoting the interquartile range. Weak but statistically significant correlations are found with stellar mass and SFR, while the dependence on sSFR is weaker.}
\label{SlopeParameters.fig}
\end{figure*}

We next examine whether the wavelength dependence of the disc scale length relative to the stellar-mass distribution depends on global galaxy properties. In particular, we consider the dependence of $h_\lambda/h_\star$ on stellar mass, SFR, and sSFR. To this end, the TSA sample is divided into bins of $M_\star$, SFR, and sSFR, and the median relation between $h_\lambda/h_\star$ and wavelength is computed separately in each bin. The resulting relations are shown in Fig.~{\ref{SlopeParameters.fig}} (top panels). For each global galaxy property, we subdivided the sample in four bins with an equal number of galaxies in each bin. The corresponding median values of each bin are indicated in the top right of the panels. In all cases, $h_\lambda/h_\star$ decreases with increasing wavelength, indicating that shorter-wavelength emission is systematically more extended than the stellar-mass distribution. A clear dependence on global galaxy properties is visible: galaxies with higher stellar masses, higher SFRs, and higher sSFRs tend to exhibit a slightly stronger wavelength dependence.

To quantify these trends on a galaxy-by-galaxy basis, we introduce a parameter $\alpha$ that describes the slope of the wavelength dependence. Specifically, for each galaxy we fit a linear relation between $\log(h_\lambda/h_\star)$ and $\log \lambda$, such that

\begin{equation}
\frac{h_\lambda}{h_\star} \propto \lambda^{-\alpha}. 
\end{equation}

\noindent The parameter $\alpha$ thus provides a compact measure of the strength of the size--wavelength relation, with larger values corresponding to a stronger decrease of scale length with increasing wavelength. As a complementary, model-independent measure of the wavelength dependence, we also define a simple parameter 
\begin{equation}
\beta = \frac{h_{\textit{u}} + h_{\textit{g}}}{h_{\text{W1}} + h_{\text{W2}}} - 1,
\end{equation}
which compares the average scale length at short (UV/optical) and long (NIR) wavelengths. We find that $\beta$ is almost perfectly correlated with $\alpha$ ($\rho = 0.999$), indicating that our results are not sensitive to the specific choice of parametrisation. In the remainder of this work, we adopt $\alpha$ as our primary measure, as it makes use of the full wavelength coverage.

The bottom panels of Fig.~\ref{SlopeParameters.fig} show $\alpha$ as a function of stellar mass, SFR, and sSFR. While significant scatter is present, weak but statistically significant correlations are found with stellar mass ($\rho = 0.19$, $p = 1.6 \times 10^{-4}$) and SFR ($\rho = 0.23$, $p = 3.8 \times 10^{-6}$), indicating that galaxies with higher masses and higher SFRs tend to exhibit a slightly stronger wavelength dependence. The correlation with sSFR is weaker ($\rho = 0.10$, $p = 0.04$). These results indicate that global galaxy properties play only a secondary role in setting the wavelength dependence of disc scale length, motivating an investigation of more local or structural drivers in the following sections.

\subsection{Disentangling stellar population gradients and dust attenuation}
\label{DustAttenuation.ssec}

\begin{figure}
\includegraphics[width=0.95\columnwidth]{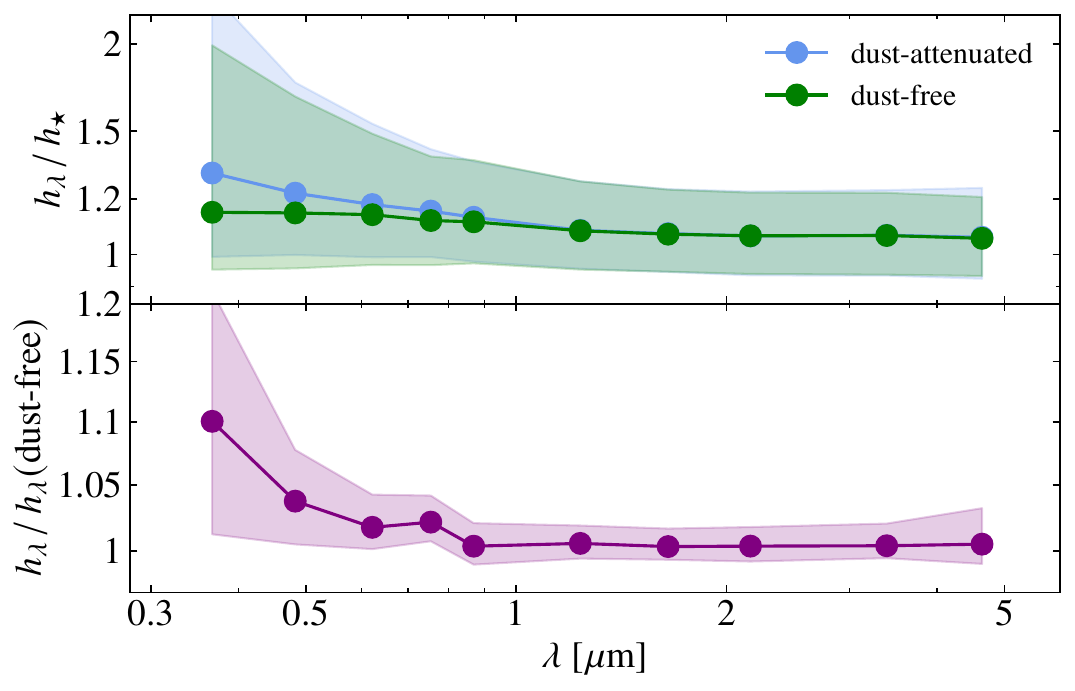}
\caption{{\it{Top panel:}} Median relation between $h_\lambda/h_\star$ and wavelength for dust-free (green curve) and dust-attenuated (blue curve) images. Shaded regions denote the interquartile range. {\it{Bottom panel:}} Ratio of dust-attenuated to dust-free scale lengths as a function of wavelength.}
\label{DustAttenuationEffect.fig}
\end{figure}

The wavelength dependence of $h_\lambda/h_\star$ established in the previous subsections may arise from two distinct effects: intrinsic radial gradients in the stellar populations that dominate the emission at different wavelengths, and wavelength-dependent dust attenuation that modifies the observed surface-brightness profiles. The TSA allows these two contributions to be separated directly by systematically comparing scale lengths measured from dust-attenuated images with those derived from dust-free images for the same galaxies.

The top panel of Fig.~{\ref{DustAttenuationEffect.fig}} shows the median relation between $h_\lambda/h_\star$ and wavelength for both dust-free and dust-attenuated images. Even in the absence of dust, a clear and monotonic decrease of $h_\lambda/h_\star$ with increasing wavelength is present. Ultraviolet and blue optical bands yield scale lengths larger than the stellar-mass scale length, while near-infrared bands approach unity. This demonstrates that a substantial wavelength dependence is already present without the effects of dust. Including dust attenuation enhances this trend at short wavelengths. In the ultraviolet and blue optical bands, the dust-attenuated scale lengths are systematically larger than their dust-free counterparts. This behaviour is consistent with centrally concentrated dust preferentially suppressing short-wavelength emission in the inner regions, thereby increasing the apparent scale length.

The lower panel of Fig.~{\ref{DustAttenuationEffect.fig}} highlights this effect more directly by showing the ratio of dust-attenuated to dust-free scale lengths as a function of wavelength. A clear excess is present at short wavelengths, reaching $\sim$10\% in the {\textit{u}} band, and decreasing steadily toward longer wavelengths. Beyond $\lambda \gtrsim 1~\mu$m, the ratio is close to unity, indicating that dust attenuation has only a minor impact in the NIR. Consistent with this behaviour, the distribution of $\alpha$ values is shifted towards higher values in the dust-attenuated case (median $\alpha = 0.086$ compared to $\alpha = 0.051$ for dust-free images), although the two distributions largely overlap. These results suggest that most of the wavelength dependence of disc scale length is already present in the absence of dust, while dust attenuation introduces an additional modulation that is primarily significant at short wavelengths.

\subsection{Age and metallicity gradients as drivers of the size--wavelength relation}
\label{AgeMetallicity.ssec}

\begin{figure*}
\includegraphics[width=\textwidth]{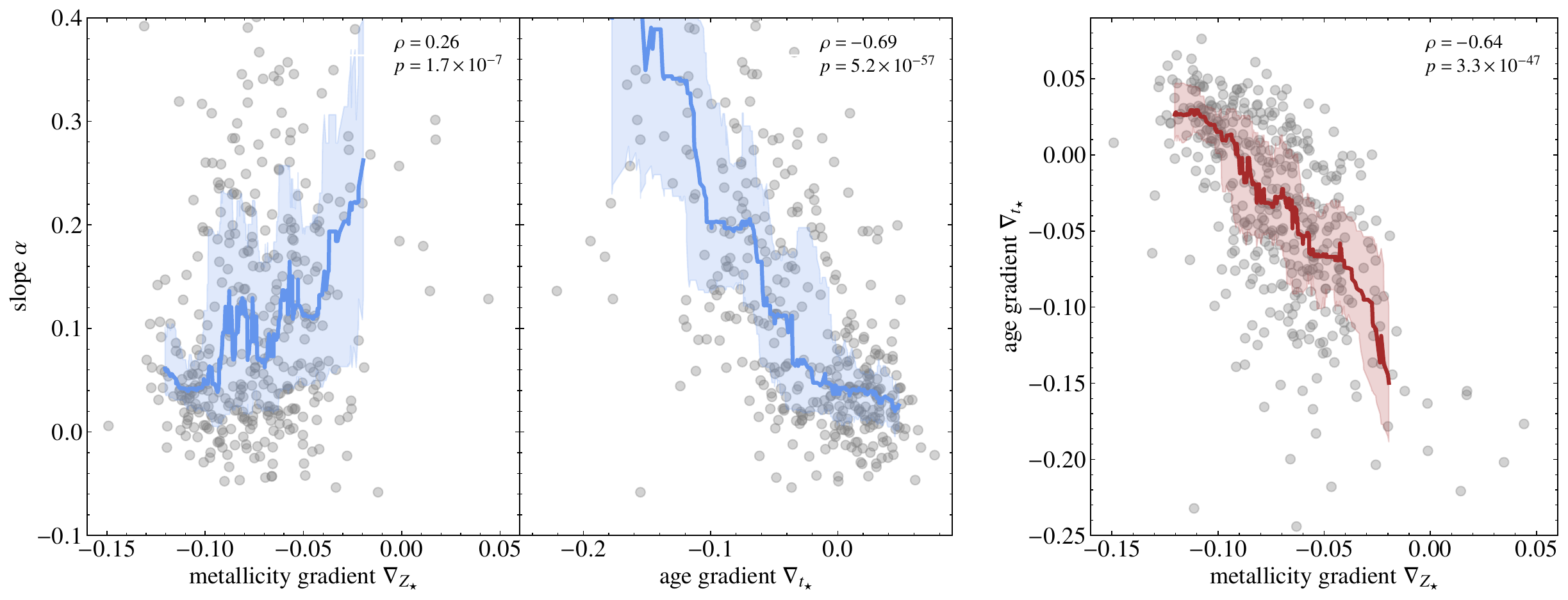}
\caption{Correlation of the slope $\alpha$ with stellar population gradients. \textit{Left:} $\alpha$ as a function of the metallicity gradient $\nabla_{Z_\star}$. \textit{Middle:} $\alpha$ as a function of the age gradient $\nabla_{t_\star}$. \textit{Right:} Relation between age and metallicity gradients. Grey points show individual galaxies, while coloured curves indicate the running median, with shaded regions denoting the interquartile range. A strong anti-correlation is observed between age and metallicity gradients. No significant correlations were found between slope and integrated age or metallicity.}
\label{PopulationGradients.fig}
\end{figure*}

To further investigate the origin of the wavelength dependence of disc scale length, we examine its relation with radial gradients in stellar population properties. In particular, we consider gradients in stellar age and stellar metallicity, which both contribute to radial colour variations in galactic discs.

For each galaxy, we quantify radial gradients in stellar age and metallicity using azimuthally averaged radial profiles constructed from the corresponding maps in the TSA. These maps have the same spatial resolution and sampling as the synthetic images and represent stellar-mass-weighted mean stellar age and metallicity \citep{Baes2024a}. The gradients are defined as the slopes of $\log t_\star$ and $\log Z_\star$ as a function of radius normalised by the stellar disc scale length, such that
\begin{gather}
\nabla_{t_\star} = \frac{{\text{d}}\log t_\star}{{\text{d}}(R/h_\star)}, 
\\
\nabla_{Z_\star} = \frac{{\text{d}}\log Z_\star}{{\text{d}}(R/h_\star)}.
\end{gather}
Here, $t_\star$ and $Z_\star$ denote the stellar-mass-weighted mean stellar age and metallicity, respectively. The gradients are obtained from linear fits to the radial profiles within the disc-dominated region of each galaxy. The definitions above express gradients in units of the stellar disc scale length, thereby providing a dimensionless measure that allows a consistent comparison across galaxies of different physical sizes.

The left and middle panels of Fig.~{\ref{PopulationGradients.fig}} display the resulting relation between the slope $\alpha$ and the metallicity and age gradients, respectively. A positive correlation is found between $\alpha$ and $\nabla_{Z_\star}$ ($\rho = 0.26$, $p = 1.7 \times 10^{-7}$), indicating that galaxies with shallower metallicity gradients tend to exhibit a stronger wavelength dependence. In comparison, the correlation with the age gradient is much stronger: a strong negative correlation is observed between $\alpha$ and $\nabla_{t_\star}$ ($\rho = -0.69$, $p = 5.2 \times 10^{-57}$), such that galaxies with more negative age gradients tend to show larger values of $\alpha$.

To further assess the relative importance of age and metallicity gradients, we perform a partial correlation analysis.\footnote{Partial correlation analysis quantifies the correlation between two variables while controlling for the effect of a third variable, thereby isolating their direct statistical association \citep{Kendall1979, Feigelson2012}. For applications of this method in astrophysics see \citet{Bait2017}, \citet{Bluck2019, Bluck2020}, or \citet{ScholzDiaz2024}.} Controlling for age gradients, the correlation between $\alpha$ and $\nabla_{Z_\star}$ becomes negative ($\rho = -0.32$, $p = 5.5 \times 10^{-11}$). In contrast, the correlation between $\alpha$ and $\nabla_{t_\star}$ remains strong when controlling for metallicity gradients ($\rho = -0.70$, $p = 2.7 \times 10^{-60}$). This indicates that the apparent correlation between $\alpha$ and metallicity gradients is not independent of age gradients.

The right panel of Fig.~{\ref{PopulationGradients.fig}} shows the relation between age and metallicity gradients. A strong anti-correlation is present ($\rho = -0.64$, $p = 3.3 \times 10^{-47}$), indicating that galaxies with steep negative age gradients typically also exhibit shallow metallicity gradients, and vice versa. This covariance between the two gradients naturally explains the behaviour seen in the partial correlation analysis.

These results demonstrate that the wavelength dependence of disc scale length is strongly linked to radial variations in stellar population properties, with distinct trends observed for age and metallicity gradients.

\section{Discussion}
\label{Discussion.sec}

\subsection{Exponential discs in TNG50: success with a tension}
\label{ExponentialDiscs.ssec}

A central structural property of late-type galaxies is the near-exponential form of their stellar discs, as established observationally for decades \citep[e.g.][]{Freeman1970, Kent1985, Pohlen2006}, often extending over several scale lengths. Despite its apparent simplicity, the physical origin of this ubiquitous structure remains debated. Proposed explanations range from angular momentum conservation during baryonic collapse within dark matter haloes \citep[e.g.][]{Fall1980, Mo1998}, to secular redistribution processes within discs, to viscous evolution of gaseous discs coupled to star formation \citep[e.g.][]{Lin1987, Wang2022a, Wang2022b}. Reproducing exponential discs in fully cosmological simulations is therefore a non-trivial requirement.

In this respect, TNG50 performs remarkably well. The {\textit{r}}-band surface brightness profiles of the star-forming galaxies in our TSA sample are close to exponential over a substantial radial range. The simulation therefore succeeds in reproducing the overall functional form that characterises observed stellar discs. This agreement is notable, as it emerges from the complex interplay of cosmological gas accretion, star formation, feedback, and internal dynamical evolution implemented in the simulation, without being explicitly imposed.

However, this success is accompanied by a clear tension. We confirm that TNG50 discs are systematically too large at fixed stellar mass, in line with previous analyses of IllustrisTNG galaxies \citep[e.g.][]{Genel2018, Wang2023}. While the exponential shape is reproduced, the radial scale of the discs is offset relative to observations. In other words, the simulation captures the form of stellar discs, but not their characteristic size.

\citet{Wang2023} argue that this size excess is likely related to insufficient angular momentum transport within the simulated gas discs. In their analysis, accreted gas retains a large fraction of its specific angular momentum after settling into the disc, leading to extended star-forming regions and, ultimately, overly large stellar discs. While we have not directly investigated the angular momentum evolution of the gas in this work, this interpretation offers a physically motivated explanation for the size offset that we confirm here.

Although angular momentum transport plays an important role in shaping galactic discs, it is unlikely to be the only process determining their sizes. Stellar feedback can regulate the removal and re-accretion of gas, thereby affecting its angular momentum and the resulting stellar distribution \citep{Crain2015, Agertz2016, Yang2024}. Indeed, galaxy sizes in IllustrisTNG have been shown to be sensitive to the adopted stellar feedback prescriptions \citep{Pillepich2018a}. AGN feedback may also influence disc growth by modifying gas accretion and star formation, although its net effect on disc size is model-dependent \citep{Grand2017, Joshi2025}.

Environmental effects may also contribute, although observational and simulation studies suggest that their influence on disc sizes is relatively modest compared to that of internal processes \citep{Weinmann2009, Du2024}. In addition, differences in disc--bulge structure and in the methods used to measure disc scale lengths may affect the comparison between simulated and observed galaxies \citep{Scannapieco2010, Joshi2025}. In particular, the different sample selection criteria and structural decomposition techniques adopted in the observational studies may contribute to the systematic offset that we find.

Taken together, these considerations suggest that the larger disc scale lengths measured in TNG50 may result from a combination of physical processes and methodological differences. Our analysis does not allow us to disentangle these contributions or identify the dominant cause of the discrepancy. A more detailed investigation, including a consistent analysis of simulated and observed galaxies, would be required to establish its physical origin.

\subsection{What drives the wavelength dependence of disc scale lengths?}

The increase of disc scale length toward shorter wavelengths is a well-established observational result \citep[e.g.][]{deJong1996b, MacArthur2003, Fathi2010, Casasola2017}. In the TSA sample, we recover the same qualitative behaviour: $h_\lambda/h_\star$ rises smoothly from the NIR to the optical and reaches its largest values in the {\textit{u}} band, the bluest band we consider in this study. The agreement with the multi-wavelength measurements of \citet{Casasola2017} confirms that the simulations reproduce both the amplitude and wavelength dependence of this trend over a broad spectral range.

A key advantage of the TSA dataset is that it allows a direct comparison between dust-attenuated and dust-free images, enabling a clean separation of the underlying physical drivers. The results presented in Sect.~{\ref{DustAttenuation.ssec}} demonstrate that most of the variation in $h_\lambda/h_\star$ is already present in the absence of dust. Dust attenuation modifies the trend in the expected sense, slightly enhancing scale lengths at short wavelengths, but its quantitative impact remains modest compared to the intrinsic variation produced by stellar population gradients.

This has important implications for the long-standing debate on the origin of size--wavelength relations in disc galaxies. Several studies have interpreted the observed decrease of scale length from the UV to the NIR primarily in terms of dust attenuation, particularly through preferential reddening of the inner regions \citep[e.g.][]{Beckman1996, Pompei1997, Mollenhoff2006}. For example, \citet{Kelvin2012}, analysing the wavelength dependence of half-light radii, concluded that dust models can largely account for the apparent size--wavelength relation in late-type disc-dominated galaxies. Other works have instead emphasised intrinsic stellar population gradients \citep[e.g.][]{deJong1996b, Cunow2004, MacArthur2004, MunozMateos2007}.

Our results clearly favour the latter interpretation, at least for the star-forming disc galaxies considered here. Even in the absence of dust, discs are intrinsically more extended at shorter wavelengths. Dust attenuation therefore acts as a secondary modulation, rather than the primary driver of the observed trend. This finding is further supported by the observational results on the SFR profiles by \citet{Casasola2017}. By deriving the SFR surface density from a combination of FUV and 22~$\mu$m emission -- thus accounting for the total (obscured + unobscured) star formation -- they found that the scale length of the total SFR is remarkably consistent with that of the stellar mass ($h_{\rm{SFR}}/h_{3.6\,\mu\rm{m}} = 0.99 \pm 0.05$). This observational evidence confirms that the current star formation effectively traces the full extent of the stellar disc, providing a physical basis for the intrinsic age gradients that drive the observed size-wavelength relation. Moreover, our analysis is qualitatively consistent with the findings of \citet{Baes2024b} for the wavelength dependence of the effective radius $R_{\text{e}}$ across the full galaxy population. However, by focusing on star-forming disc galaxies and explicitly measuring disc scale lengths rather than global half-light radii, our analysis isolates the effect at the level of the disc component itself.

The origin of this intrinsic wavelength dependence can be further understood in terms of the radial structure of stellar populations. The analysis presented in Sect.~{\ref{AgeMetallicity.ssec}} shows that variations in $\alpha$ are strongly linked to radial age gradients, while the dependence on metallicity gradients is significantly weaker once their mutual correlation is taken into account. This indicates that the size--wavelength relation is primarily driven by the spatial distribution of stellar ages: galaxies in which younger stellar populations are more extended naturally exhibit larger scale lengths at shorter wavelengths.

In this picture, the wavelength dependence of disc scale length directly reflects the inside-out growth of galactic discs. In such a scenario, star formation proceeds progressively at larger radii over cosmic time, leading to younger stellar populations being more spatially extended than older ones. Observational evidence for inside-out growth has been found in resolved stellar population studies and multi-wavelength analyses of nearby galaxies \citep[e.g.][]{MunozMateos2007, GonzalezDelgado2015, Lian2017, Pessa2023}, while cosmological simulations similarly predict the build-up of discs through the gradual outward growth of the star formation \citep[e.g.][]{Clauwens2017, Genel2018, Trayford2019}. The resulting age gradients imprint themselves on the light distribution in a wavelength-dependent manner, producing the observed increase of scale length toward shorter wavelengths. While metallicity gradients are also present, their role appears to be secondary and largely coupled to the age structure of the disc.

\subsection{Caveats, limitations, and prospects}

While the results presented in this work are robust at the population level, several caveats should be kept in mind.

First, the identification of the disc-dominated radial range relies on visual inspection and quality assessment of exponential fits. Although multiple independent assessors were used and a clear quality criterion was imposed, the procedure remains partly subjective. This is difficult to avoid in practice, as automated fitting over fixed radial intervals can be strongly biased by bulges, bars, or other central components. Nevertheless, alternative choices of the fitting range would mainly affect individual galaxies rather than the global trends discussed here.

Second, our analysis is based on azimuthally averaged surface brightness profiles and one-dimensional exponential fits. This approach intentionally suppresses non-axisymmetric features such as bars, spiral structure, or lopsidedness. In real galaxies, such structures can influence fitted scale lengths, especially when two-dimensional decompositions are performed \citep[e.g.][]{Gadotti2009}. Our methodology is designed to isolate the large-scale exponential disc component, but it inevitably simplifies the internal complexity of discs.

Third, although the TSA galaxies were selected to be close to face-on, they are not perfectly so. Residual inclination effects and projection-induced asymmetries may introduce small systematic uncertainties in the measured scale lengths. These effects are expected to be modest given the selection criteria, but they cannot be completely eliminated in the current dataset.

A further limitation concerns the wavelength coverage. In the present work, the analysis extends from the {\textit{u}} band to WISE W2. The absence of GALEX FUV/NUV bands and of mid- and far-infrared dust emission limits the ability to probe the full extent of the wavelength dependence of disc structure. In particular, comparisons at very short wavelengths are restricted, as the current UV mock images in TSA DR1 suffer from stochastic sampling of young stellar particles, leading to artificially clumpy morphologies. At long wavelengths, the lack of thermal dust re-emission modelling prevents a direct structural comparison in the infrared and submm regime.

Several of these limitations have been addressed in the recently released TSA DR2 \citep{Baes2026}. In addition to the existing random orientations, DR2 provides dedicated face-on and edge-on projections, reducing projection-related uncertainties. The treatment of young stellar populations has also been improved, enabling more realistic UV imaging, while the inclusion of thermal dust emission extends the wavelength coverage to the submm regime. These improvements open the way for a more comprehensive investigation of the wavelength dependence of disc scale lengths and its physical drivers. The present analysis is based on the DR1 data products, and extending it to the improved DR2 dataset would be a natural direction for future work.

Another caveat concerns the adopted dust recipe in the TSA radiative transfer modelling. In the current implementation, the dust distribution is linked to the simulated metal distribution and is therefore determined by the underlying assumptions used to construct the SKIRT input models. Alternative dust recipes could modify the detailed attenuation pattern within galaxies and consequently affect the measured scale lengths, particularly at optical wavelengths where dust attenuation is strongest. For example, a more centrally concentrated dust distribution would increase central obscuration and tend to flatten the observed surface-brightness profiles, whereas a smoother or more extended dust configuration would produce weaker attenuation effects. Although our comparison between dust-free and dust-attenuated simulations demonstrates that dust is not the primary driver of the size–wavelength relation for the dust prescription adopted in this work, the quantitative contribution attributed to dust may depend on the adopted dust recipe.

A further caveat concerns the potential contribution of AGN to the observed disc scale lengths. As discussed in Sect.~\ref{SampleSelection.ssec}, direct AGN emission is not included in the TSA radiative transfer calculations. However, since AGN emission is generally concentrated in the central regions of galaxies, which are excluded from our exponential disc fits, we do not expect its omission to significantly affect the measured disc scale lengths. In the observational comparison samples, some residual AGN contamination cannot be entirely excluded, particularly if light from a luminous AGN extends into the fitted disc region through the wings of the instrumental PSF. This effect is expected to be limited for the low-luminosity AGN present in approximately half of the DustPedia sample \citep{Casasola2017}. Importantly, the absence of direct AGN emission in the TSA images does not imply that AGN activity has no influence on the simulated galaxies. AGN feedback is included in the TNG galaxy formation model and can affect the intrinsic structure and evolution of galactic discs, as discussed in Sect.~\ref{ExponentialDiscs.ssec}. In future work, direct AGN emission could be incorporated into the SKIRT post-processing \citep{Stalevski2012,Stalevski2016} to investigate its impact on measured disc scale lengths more quantitatively.

Finally, the present study is restricted to $z=0$, reflecting the current scope of TSA release. The cosmic evolution of galaxy sizes has been an important constraint on galaxy evolution models for more than a decade \citep[e.g.][]{vanderWel2014, Lange2015, Mowla2019, Suess2019, deGraaff2022}, and has become a hot topic in the current {\it{JWST}} era \citep[e.g.][]{Ito2024, Ward2024, Martorano2024, Miller2025, Song2026}. The dependence of galaxy size on wavelength is recognised as an important ingredient in interpreting this evolutionary trend \citep{vanderWel2024, Jia2024, Xin2026}. Several recent observational studies have highlighted that the size--wavelength relation may vary with redshift, potentially reflecting changes in stellar population gradients, dust content, and the spatial distribution of star formation. Extending the present analysis to higher redshift would therefore provide valuable insight into how the physical drivers identified here evolve over cosmic time. Encouragingly, multi-wavelength radiative transfer modelling of TNG50 galaxies at higher redshift has already been explored in recent work \citep[e.g.][]{Costantin2023}, demonstrating the feasibility of such studies. A future TSA extension including multiple redshift snapshots would enable a systematic investigation of the redshift evolution of disc scale lengths and their wavelength dependence within a fully self-consistent framework.

\section{Summary}
\label{Summary.sec}

In this work we have used the TNG50--SKIRT Atlas to investigate the structure of star-forming disc galaxies at $z=0$, focusing on exponential scale lengths and their dependence on wavelength. Our main findings can be summarised as follows.
\begin{itemize}

\item The stellar discs of TNG50 galaxies are close to exponential over a substantial radial range. In this sense, the simulation successfully reproduces one of the defining structural properties of late-type galaxies. However, at fixed stellar mass, the discs are systematically larger than observed, indicating a persistent offset in the normalisation of the size--mass relation, consistent with previous studies \citep{Wang2023}.

\item We have quantified the wavelength dependence of disc scale lengths using the dimensionless ratio $h_\lambda/h_\star$. The TSA galaxies exhibit a clear and monotonic increase of scale length toward shorter wavelengths, in good agreement with observational results over the optical-to-near-infrared range. This demonstrates that the simulation reproduces the overall form of the observed size--wavelength relation.

\item By exploiting the availability of both dust-free and dust-attenuated mock images, we have shown that this wavelength dependence is primarily driven by intrinsic stellar population gradients. More specifically, radial gradients in stellar age are the dominant contributor to the effect, while metallicity gradients play a secondary role. Dust attenuation enhances the wavelength dependence at short wavelengths, but does not set its overall amplitude. This establishes a clear physical interpretation of the size--wavelength relation, in which the spatial distribution of stellar populations, shaped by the star formation history of the disc, is the principal driver.

\item Taken together, these results show that cosmological simulations such as TNG50 can reproduce both the exponential structure of stellar discs and the observed wavelength dependence of disc scale lengths, while also providing insight into its physical origin. At the same time, the systematic offset in absolute disc scale length highlights the need for further investigation of both the physical processes governing disc formation and the methodologies used to compare simulated and observed galaxies. The recently released TSA DR2 \citep{Baes2026} provides improved modelling and extended wavelength coverage, enabling more comprehensive investigations of disc structure. Future extensions to higher redshift will allow these structural diagnostics to be applied across cosmic time, offering more stringent tests of galaxy formation models and the evolution of stellar population gradients in galactic discs.
\end{itemize}

\begin{acknowledgements}
We thank the anonymous referee for their constructive comments and suggestions, which helped us improve the clarity and presentation of the manuscript. SBT and ATE appreciate financial support from the NASCERE project, a bilateral cooperation program between Jimma University (Ethiopia) and Ghent University (Belgium). MB and MM acknowledge financial support from the Flemish Fund for Scientific Research (FWO-Vlaanderen) through the research project G030319N. MB and AN acknowledge support from the Belgian Science Policy Office (BELSPO) through the PRODEX project ``Belgian Euclid Science Exploitation (BESE)'' (No. 4000143202).
\end{acknowledgements}

\bibliography{mybib_nameyear}

\end{document}